\documentclass[journal=jctcce, 
manuscript=article]{achemso}

\usepackage{amsmath}
\usepackage{amssymb}
\usepackage{graphicx}
\usepackage{dcolumn}
\usepackage{bm}
\usepackage{tabularx}
\usepackage{xcolor}
\usepackage{tikz}
\usetikzlibrary{quantikz}
\usepackage[version=4]{mhchem}
\usepackage{physics}
\usepackage{braket}
\usepackage{soul}
\usepackage{nicefrac}
\usepackage{xr-hyper}
\usepackage{hyperref}
\usepackage{natbib}
\usepackage{subcaption}
\usepackage{longtable} 
\usepackage{notes2bib}
\usepackage{url}
\usepackage{array}
\usepackage{multirow}
\usepackage{marginnote}

\title{Quantum simulations of Fermionic Hamiltonians with efficient encoding and ansatz schemes}

\author{Benchen Huang}
\affiliation{Department of Chemistry, University of Chicago, Chicago, IL 60637, USA}%

\author{Nan Sheng}
\affiliation{Department of Chemistry, University of Chicago, Chicago, IL 60637, USA}%

\author{Marco Govoni}
\affiliation
{Pritzker School of Molecular Engineering, University of Chicago, Chicago, IL 60637, USA.}
\alsoaffiliation
{Materials Science Division and Center for Molecular Engineering, Argonne National Laboratory, Lemont, IL 60439, USA.}
\email{mgovoni@anl.gov}

\author{Giulia Galli}
\affiliation
{Pritzker School of Molecular Engineering, University of Chicago, Chicago, IL 60637, USA.}
\alsoaffiliation
{Department of Chemistry, University of Chicago, Chicago, IL 60637, USA.}
\alsoaffiliation[MSD]
{Materials Science Division and Center for Molecular Engineering, Argonne National Laboratory, Lemont, IL 60439, USA.}
\email{gagalli@uchicago.edu}

\begin{document}

\begin{abstract}
We propose a computational protocol for quantum simulations of Fermionic Hamiltonians on a quantum computer, enabling calculations on spin defect systems which were previously not feasible using conventional encodings and unitary coupled-cluster ansatz of variational quantum eigensolvers. We combine a qubit-efficient encoding scheme mapping Slater determinants onto qubits with a modified qubit-coupled cluster ansatz and noise-mitigation techniques. Our strategy leads to a substantial improvement in the scaling of circuit gate counts and in the number of required qubits, and to a decrease in the number of required variational parameters, thus increasing the resilience to noise. We present results for spin defects of interest for quantum technologies, going beyond minimum models for the negatively charged nitrogen vacancy center in diamond and the double vacancy in 4H silicon carbide (4H-SiC) and tackling a defect as complex as negatively charged silicon vacancy in 4H-SiC for the first time.
\end{abstract}

\maketitle


\section{Introduction} \label{introduction}

Obtaining accurate solutions of the electronic structure of many-body systems is a major challenge in computational science, and an important endeavor that may benefit problems in several fields, ranging from catalysis~\cite{bell2011quantum, xu2018theoretical, hammes2021integration} and drug discovery~\cite{jorgensen2004many} to quantum technologies~\cite{ladd2010quantum, degen2017quantum}. In addition to steady efforts in the development of algorithms to solve the electronic structure problem on classical computers, research into the use of quantum computers to solve the time independent Schr\"odinger equation has been flourishing in the past decades~\cite{aspuru2005simulated, peruzzo2014variational, somma2019quantum, parrish2019quantum, ge2019faster, motta2020determining, lin2022heisenberg, huggins2022unbiasing}. The motivation behind this trend is the promise that a fault-tolerant quantum computer may be able to solve the electronic structure problem for  many-body systems~\cite{cao2019quantum} in polynomial time, for example using a quantum phase estimation (QPE)~\cite{kitaev1995quantum, abrams1997simulation, abrams1999quantum, aspuru2005simulated} algorithm. The latter is a probabilistic method to obtain the eigenstate of a unitary operator that assumes that the initial state of a given system, prepared on a quantum computer, has a non vanishing overlap with the target state.

The possibility of reaching exponential quantum advantage for quantum chemistry problems remains controversial~\cite{lee2022there}. However, it is interesting to explore whether quantum computers may in fact turn out to be advantageous over classical ones, even in the absence of exact polynomial scaling, and in particular whether even today’s noisy intermediate scale quantum (NISQ) platforms may be utilized for interesting problems. Recent efforts~\cite{huggins2022unbiasing, xu2022quantum, zhang2022quantum} to incorporate quantum computations into quantum Monte Carlo methods~\cite{motta2018ab} suggest new route for such benefits to be achieved, in practice, even with noisy hardware. Specifically, Ref~\cite{huggins2022unbiasing} reported a calculation of the atomization energy of the strongly correlated square H$_4$ molecule, using a quantum-classical hybrid quantum Monte Carlo method on the Sycamore quantum processor~\cite{arute2019quantum}, which achieved accuracy that is competitive with state-of-the-art classical methods. The algorithm relies on the preparation of a so called \textit{a priori} quantum trial state on the quantum hardware, which is considered as an approximation to the target ground state. Therefore it appears that one strategy to obtain computational advantage on both NISQ and fault-tolerant quantum devices, relies on the efficient preparation of an accurate initial state; this strategy has been explored for both molecular~\cite{lanyon2010towards, peruzzo2014variational, o2016scalable, hempel2018quantum, kandala2017hardware, kandala2019error, smart2019quantum, google2020hartree, yeter2020practical, smart2021quantum, smart2022resolving, eddins2022doubling, huang2022leveraging} and condensed systems~\cite{ma2020quantum, cerasoli2020quantum, sherbert2021systematic, yamamoto2022quantum, huang2022simulating}.

An appealing and popular protocol to obtain the ground state of Fermionic systems is that of writing the Hamiltonian in second quantization and using a variational quantum eigensolver (VQE)~\cite{peruzzo2014variational, mcclean2016theory}. This algorithm parameterizes the many-body wavefunction through a quantum circuit, and the energy is measured on a noisy hardware. Upon  optimization of the parameters on classical hardware, one obtains a variational upper bound on the ground state energy. The efficiency and reliability of VQE depend on the number of available qubits on the quantum hardware, on the qubit coherence time and usually VQE faces optimization challenges due to the hardware noise. However, despite theses challenges, this algorithm has been successfully applied to study systems with up to 12 electrons~\cite{google2020hartree}.

In a recent paper, we utilized VQE to solve the electronic structure of the minimum model of realistic solid state systems with strongly correlated states, and we carried out calculations on a quantum computer. In particular we considered spin-defects in solids, i.e. the negatively charged nitrogen vacancy center (\ce{NV^-}) in diamond and the neutral di-vacancy (\ce{VV^0}) in 4H-SiC~\cite{huang2022simulating}, which are of interest for quantum information applications~\cite{weber2010quantum, wolfowicz2021quantum}, including quantum sensing~\cite{hsieh2019imaging}, communication~\cite{anderson2022five} and bioimaging~\cite{shi2015single}. Although we obtained encouraging results,  we also identified several problems awaiting for more efficient and accurate solutions. For example, the so called unphysical state problem~\cite{sawaya2016error}, caused by an imperfect  conservation of the number of particles on a noisy hardware,  leads to values of the energy that lie below the exact classical reference value. We solved this problem by post selecting~\cite{huggins2021efficient} the measured values of the energy and considering only those corresponding to the  correct number of particles. The combination of post-selection and zero-noise extrapolation (ZNE) techniques~\cite{li2017efficient, temme2017error, endo2018practical, endo2021hybrid} led us to solve the electronic structure of realistic spin-defects. However, we could do so only for minimum models, as the ansatz circuit used in VQE usually leads to a large gate count and hence calculations are hard to scale.

Here we propose a computational strategy leading to an improved scaling with gate counts of VQE optimizations, thus enabling electronic structure calculations of complex spin-defects previously not feasible with conventional VQE algorithms. In particular, we combine a qubit-efficient encoding (QEE) scheme~\cite{shee2022qubit} with a modified qubit-coupled cluster (QCC) ansatz~\cite{ryabinkin2018qubit} and noise-mitigation techniques. Such a protocol leads to a  substantial decrease in the number of required variational parameters in VQE calculations, thus increasing the resilience to noise and enabling calculations of spin defects beyond the minimum model~\cite{huang2022simulating}.  The rest of the paper is organized as follows. In Section~\ref{Methods} we discuss the quantum algorithms adopted to solve the electronic structure of systems whose parametrized Hamiltonian is expressed in second quantization. In Section~\ref{Results}, we present calculations on a real quantum computer of three spin defect systems, i.e. \ce{NV^-} in diamond, \ce{VV^0} and a new defect--negatively charged silicon vacancy (\ce{V^-_{Si}}) in 4H-SiC, which for some applications~\cite{kraus2014magnetic,soykal2016silicon} is a promising alternative to NV centers. Section~\ref{Conclusions} concludes our work with a summary and outlook.

\clearpage

\section{Methods} \label{Methods}

\begin{figure}[hbt!]
    \centering
    \includegraphics[width=0.4\textwidth]{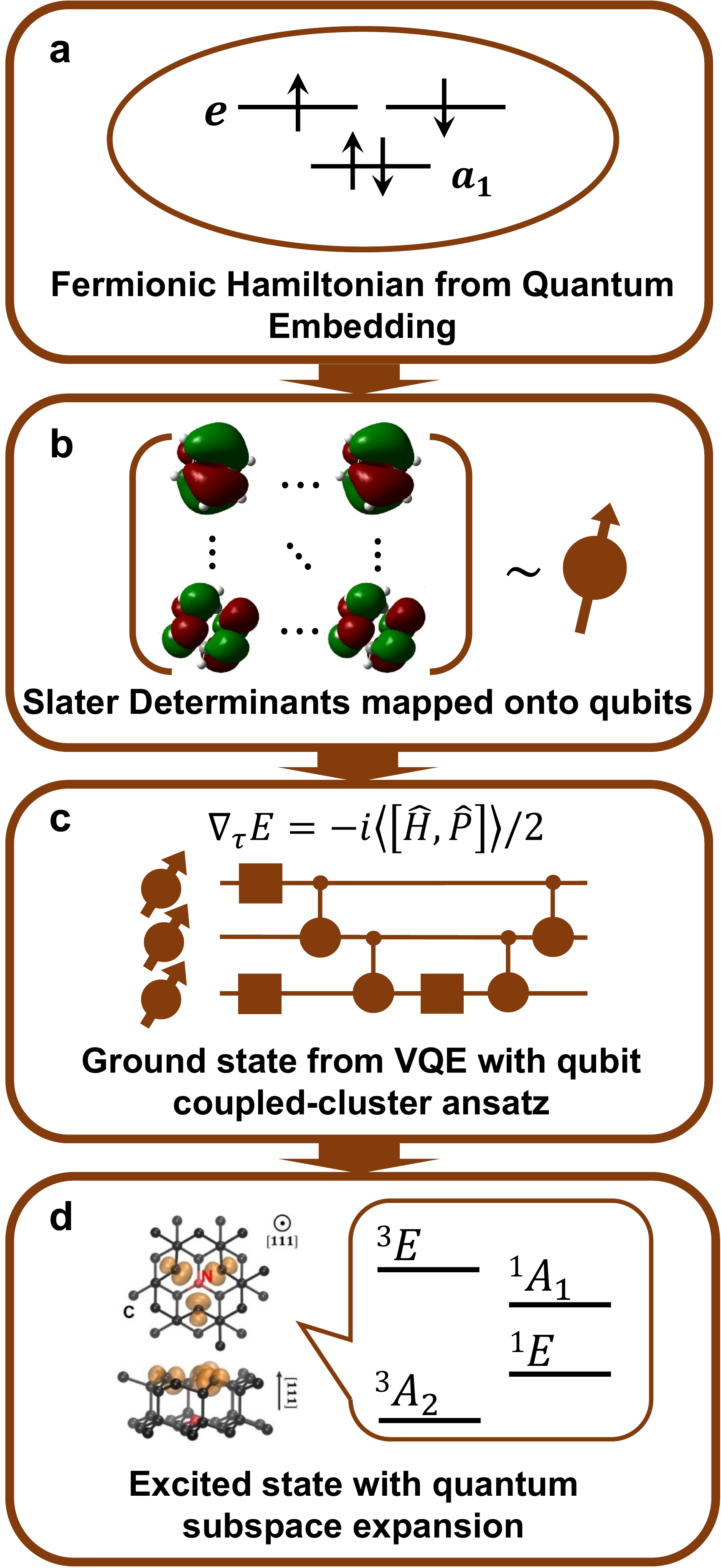}
    \caption{Workflow used to simulate the ground and excited state energies of spin defects on a quantum computer. \textbf{a} The effective Hamiltonian in second quantization describing the electronic structure of spin defects is obtained from a quantum defect embedding theory (QDET), see Sec~\ref{results:classical} for detail. \textbf{b} The Slater determinants are mapped onto qubits using a qubit-efficient encoding scheme, where the molecular orbitals represent a Slater Determinant. \textbf{c} The ground state of the effective Hamiltonian is obtained using a variational quantum eigensolver (VQE) and a qubit coupled-cluster (QCC) ansatz. \textbf{d} The excited states of the effective Hamiltonian are obtained using a quantum subspace expansion (QSE) algorithm.}
    \label{fig:workflow}
\end{figure}

The workflow adopted here to obtain the ground and excited states of a Fermionic Hamiltonian $\hat{H}_\mathrm{elec}$ on a quantum computer is summarized in Fig.~\ref{fig:workflow} and consists of the following steps: (i) define a Fermionic Hamiltonian using a quantum defect embedding theory (QDET), (ii) derive a qubit Hamiltonian by mapping selected electronic configurations (Slater determinants) of the Fermionic Hamiltonian onto qubits, (iii) compute the ground state energy of the qubit Hamiltonian using VQE, (iv) compute the excited states using the quantum subspace expansion (QSE) algorithm.
The qubit Hamiltonian $\hat{H}_{\mathrm{q}} = \sum_i g_i \hat{P}_i$ contains coefficients $g_i$ obtained from the one- and two- body terms of $\hat{H}_\mathrm{elec}$ multiplied by Pauli strings, i.e., $\hat{P}_i \in \{I, X, Y, Z\}^{\otimes N_q}$, where $N_q$ is the number of qubits and $I$, $X$, $Y$, $Z$ are Pauli operators. When using a VQE algorithm, an ansatz circuit, usually a parametrized unitary operator $\hat{U}(\Vec{\theta})$, is defined and applied to a chosen initial state $\ket{\Psi_0}$. Finally, the ground state energy $E_g$ is variationally obtained by optimizing the parameters $\Vec{\theta}$ of the ansatz such that $E_g = \min_{\Vec{\theta}}\langle \Psi_0 |\hat{U}^{\dagger}(\Vec{\theta})\hat{H}\hat{U}(\Vec{\theta})|\Psi_0\rangle$. As mentioned above, excited states are obtained with the QSE algorithm.

For a detailed discussion of the derivation of a Fermionic Hamiltonian describing spin defects in solids using QDET we refer the reader to Ref.~\cite{ma2020quantum, ma2021quantum, sheng2022green, vorwerk2022quantum}. Below we discuss in details steps (ii--iv: see panels \textbf{b}-\textbf{d} of Fig.~\ref{fig:workflow}). 

\subsection{Qubit Efficient Encoding for Fermionic Mapping} \label{encoding}

The Fermion to qubit encoding is an isometry $\mathcal{E}: \mathcal{H}_{\mathrm{elec}} \to \mathcal{H}_{\mathrm{q}}$~\cite{bravyi2017tapering}, where $\mathcal{H}_{\mathrm{elec}}$ and $\mathcal{H}_{\mathrm{q}}$ represent the physical and qubit Hilbert space spanned by the eigenvectors of $\hat{H}_\mathrm{elec}$ and $\hat{H}_\mathrm{q}$, respectively. Commonly used encoding schemes such as the Jordan–Wigner (JW)~\cite{jordan1993algebraic}, Bravyi–Kitaev (BK)~\cite{seeley2012bravyi} and parity encoding methods~\cite{bravyi2017tapering} require $N_q=N$ qubits for a system with $N$ spin-orbitals, and generate a $2^N$-dimensional $\mathcal{H}_{\mathrm{q}}$. However, our goal is to compute the eigenvectors and eigenvalues of $\hat{H}_{\mathrm{elec}}$ subject to specific physical constraint on the number of electrons in the two spin channels $(m_{\uparrow},m_{\downarrow})$. In practice this constraint can be enforced by restricting the solutions of the VQE or QSE algorithms to a subspace of the qubit Hilbert space with dimension $Q=\begin{psmallmatrix} \nicefrac{N}{2} \\ {m_{\uparrow}} \end{psmallmatrix} \times \begin{psmallmatrix}\nicefrac{N}{2} \\ {m_{\downarrow}} \end{psmallmatrix}<2^N$. However, the JW and BK encoding maps do not enforce such a physical constraint and thus lead to a qubit Hilbert space that is larger than the physical one~\cite{sawaya2016error}, e.g., the former contains all Fock states, some corresponding to a number of electrons different from those of the physical system. In principle, on a fault-tolerant computer the VQE algorithm should preserve the initial number of electrons throughout the optimization process; however, the noise present in NISQ devices does not guarantee the preservation of the physical constraints~\cite{elfving2021simulating}, leading to errors in ground state energies that in Ref.~\cite{huang2022simulating} we have mitigated with a post-selection procedure. Note that other symmetry constraints, e.g., point group symmetry~\cite{setia2020reducing} could also be taken into account when choosing relevant Slater determinants, which would be interesting to explore in future works.

Here we adopt instead the QEE scheme~\cite{shee2022qubit}, a compact Fermion to qubit encoding map~\cite{kirby2021quantum, kirby2022second, chamaki2022compact} that by definition excludes from the qubit Hilbert space all Fock states with nonphysical number of electrons, leading to a robust solution of the unphysical state problem. The QEE encoding has also the benefit of requiring a smaller number of qubits than the conventional encoding maps. The use of QEE has already been shown to be beneficial on quantum hardware~\cite{shee2022qubit} for molecules such as H$_2$ and LiH; here we show that its use is crucial in the case of spin-defects, where the number of qubits required to go beyond minimum models by conventional encodings would be impractical on NISQ devices.

In the QEE scheme, one pre-selects all the  electronic configurations $\mathcal{F} = \{|\mathbf{f}\rangle_i \big| |\mathbf{f}\rangle_i \in \mathcal{H}_{\mathrm{elec}}\}$ that satisfy the required set of physical constraints, e.g., fixed number of particles and fixed spin projection $\hat{S}_z$. The implementation of the QEE scheme requires $N_q = \Big\lceil\mathrm{log}_2 Q\Big\rceil < N$ qubits. Using the QEE isometry, configurations in $\mathcal{F}$ are mapped to $\mathcal{Q} = \{|0\rangle_q , |1\rangle_q\}^{\otimes N_q}$, the computational basis states of a $N_q$-qubit system. To reduce the state preparation error~\cite{shee2022qubit}, a good practice in defining the QEE isometry $\mathcal{E}$ is to first sort both $\mathcal{F}$ and $\mathcal{Q}$ in ascending order according to the electronic energy of $|\mathbf{f}\rangle_i$ and the decimal number associated to the binary string representing the qubit state. By doing so, a correspondence $\mathcal{E}|\mathbf{f}\rangle_i = |\mathbf{q}\rangle_i$ is established. We note that in general the size $Q$ is not necessarily a power of 2. To fit the requirements of quantum circuits, unphysical states may therefore be included in QEE so as to build a Hilbert space with a size that is a power of 2. In this case, post-selection of measurement results may be helpful to exclude results involving unphysical states.

In common Fermionic-to-qubit encoding schemes, there is a one to one correspondence of both creation and annihilation operators ($\hat{a}^{\dagger}_p$, $\hat{a}_q$) with qubit operators. In QEE, where by definition only states with fixed number of particles are considered, there is a one to one correspondence between the excitation operator $\hat{E}_{pq} \equiv \hat{a}^{\dagger}_p \hat{a}_q$ and a qubit operator $\hat{\widetilde{E}}_{pq}$, where the excitation operators are first rewritten as a sum of projection of Slater determinants $\ket{\textbf{f}_i}\bra{\textbf{f}_j}$ and then transformed into qubit space through four entry operators: $\frac{1}{2}(X+iY), \frac{1}{2}(X-iY), \frac{1}{2}(I-Z), \frac{1}{2}(I+Z)$, see Ref~\cite{shee2022qubit} for detail. The qubit Hamiltonian can then be constructed using $\hat{\widetilde{E}}_{pq}$. For a generic Hamiltonian, where the projection operators would lead to a linear combination of up to an exponential number of Pauli operators, and in principle one needs to consider an exponentially large number of determinants, no quantum advantage would be achieved. QEE therefore should be considered as an intermediate solution for NISQ hardware. However, we note that both the size of the qubit Hamiltonian and the total number of Slater determinants scale polynomially as a function of $N$ for the systems considered in our study, as we explain in the Section~\ref{results:classical}.

\subsection{Qubit Coupled-Cluster Ansatz for Variational Quantum Eigensolvers} \label{ansatz}

After constructing a qubit Hamiltonian using the QEE encoding, we discuss the choice of the wavefunction ansatz. One popular ansatz used in the literature is the unitary coupled-cluster (UCC) ansatz~\cite{peruzzo2014variational, romero2018strategies, grimsley2019trotterized}, inspired by coupled-cluster theory~\cite{helgaker2014molecular}. Such an ansatz can yield accurate results for many-body systems, but it leads to calculations suffering from poor scaling as a function of the number of gates, due to the inclusion of all possible electronic excitations. A typical implementation of the UCC ansatz on quantum computers leads to the following expression in terms of Pauli strings (entanglers) $\hat{P}_k$:
\begin{equation}
    \hat{U}_{\mathrm{UCC}} = \prod_{k} \hat{U}_k = \prod_{k} e^{-i\theta_k \hat{P}_k/2}, \;\;\hat{P}_k \in \{I, X, Y, Z\}^{\otimes N_q}.
    \label{eq:UCC}
\end{equation}
The number of entanglers required when using the UCC ansatz may be large even for intermediate scale systems with $4\sim 6$ electrons. The QCC ansatz~\cite{ryabinkin2018qubit} bypasses the formulation of the ansatz in physical space, and instead directly implements Eq.~\ref{eq:UCC} in the qubit space. In particular, the QCC method proposed in Ref.~\cite{ryabinkin2018qubit} implements a screening process to select and retain the entanglers that contribute the most to the evaluation of the energy.

In QCC, the variation of the energy induced by each entangler is evaluated by expanding the energy to second order in the parameter $\theta_k$:
\begin{equation}
    \delta E[\theta_k; \hat{P}_k] = E[\theta_k; \hat{P}_k] - E_0 \approx \theta_k \frac{d E[\theta_k; \hat{P}_k]}{d\theta_k} \bigg|_{\theta_k=0} + \frac{\theta_k^2}{2} \frac{d^2 E[\theta_k; \hat{P}_k]}{d \theta_k^2} \bigg|_{\theta_k=0}, \label{eq:taylor}
\end{equation}
where $E[\theta_k; \hat{P}_k] = \langle \Psi_0 |\hat{U}_k^{\dagger} \hat{H}\hat{U}_k |\Psi_0\rangle$ and $E_0 = \langle \Psi_0 |\hat{H}|\Psi_0\rangle$. The first derivative in Eq.~\ref{eq:taylor} can be efficiently computed through quantum measurements as
\begin{equation}
    \frac{d E[\theta_k; \hat{P}_k]}{d\theta_k} \bigg|_{\theta_k=0} = \bigg\langle \Psi_0\bigg|-\frac{i}{2}\left[\hat{H}, \hat{P}_k\right]\bigg|\Psi_0 \bigg\rangle, \label{eq:first_order}
\end{equation}
which results from the similarity-transformed Hamiltonian being in closed form~\cite{lang2020unitary} 
\begin{equation}
    \hat{U}_k^{\dagger}\hat{H}\hat{U}_k = \hat{H} - i\frac{\sin\theta_k}{2}\left[\hat{H}, \hat{P}_k\right] + \frac{1}{2}\left(1-\cos\theta_k\right)\hat{P}_k\left[\hat{H}, \hat{P}_k\right].
\end{equation}
The expression of the second order derivative can be found in Ref.~\cite{ryabinkin2018qubit}. 

The implementation of the QCC method proceeds by ranking the entanglers according to the magnitude of their first-order derivative and sign of the second-order derivative, and by considering only the entanglers with highest rank. This amounts to screening the value of the first and second derivative of the energy for each of the $\sim 4^N$ entanglers and choosing those with values of the first derivatives substantially different from zero or second derivatives substantially smaller than zero. By using a qubit basis state as $\ket{\Psi_0}$ we can reduce the dependency of the total number of $\hat{P}_k$ with respect to the number of qubits from exponential to polynomial. The reduction is achieved by grouping the terms in the Hamiltonian and performing the pre-screening within each group; see Ref.~\cite{ryabinkin2020iterative} for detail. In practice, the second derivatives could also be neglected to decrease computational cost~\cite{ryabinkin2020iterative}, as we did in this work.

\begin{figure}[t!]
    \centering
    \begin{quantikz}[row sep=0.4cm, column sep=0.5cm]%
    q_0 & \gate{R_z}\gategroup[4,steps=2,style={dashed, rounded corners,fill=blue!20, inner xsep=2pt}, background, label style={label position=below, anchor=north, yshift=-0.2cm}]{{\sc QMF}} & \gate{R_x} & \gate{R_x(-\frac{\pi}{2})} & \ctrl{1} & \qw & \qw & \qw & \qw & \qw & \ctrl{1} & \gate{R_x(\frac{\pi}{2})} & \qw\\
    q_1 & \gate{R_z} & \gate{R_x} & \gate{H} & \targ{} & \ctrl{1} & \qw & \qw & \qw & \ctrl{1} & \targ{} & \gate{H} & \qw \\
    q_2 & \gate{R_z} & \gate{R_x} & \gate{H} & \qw & \targ{} & \ctrl{1} & \qw & \ctrl{1} & \targ{} & \qw & \gate{H} & \qw\\
    q_3 & \gate{R_z} & \gate{R_x} & \gate{H} & \qw & \qw & \targ{} & \gate{R_z} &\targ{} & \qw & \qw & \gate{H} & \qw
    \end{quantikz}\\
    \begin{quantikz}[row sep=0.4cm, column sep=0.5cm]%
    q_0 & \gate{R_y} & \gate{R_x(-\frac{\pi}{2})} & \ctrl{1} & \qw & \qw & \qw & \qw & \qw & \ctrl{1} & \gate{R_x(\frac{\pi}{2})} & \qw\\
    q_1 & \qw & \gate{H} & \targ{} & \ctrl{1} & \qw & \qw & \qw & \ctrl{1} & \targ{} & \gate{H} & \qw \\
    q_2 & \gate{R_y} & \gate{H} & \qw & \targ{} & \ctrl{1} & \qw & \ctrl{1} & \targ{} & \qw & \gate{H} & \qw\\
    q_3 & \qw & \gate{H} & \qw & \qw & \targ{} & \gate{R_z} &\targ{} & \qw & \qw & \gate{H} & \qw
    \end{quantikz}
    \caption{The upper panel shows a representative quantum circuit representing the qubit coupled-cluster ansatz with 4 qubits. The box circled by the dashed line shows the qubit mean-field (QMF) part of the circuit, which enables the construction of any product states. The circuit component following the QMF part enbales the construction of the exponential of entangler $XXXY$, and it is built with the CNOT gate ladders. The lower panel shows a circuit representative of the modified ansatz, where the three entanglers are $IIIY, IYII, XXXY$, as defined in the pre-screening process.}
    \label{fig:qcc}
\end{figure}
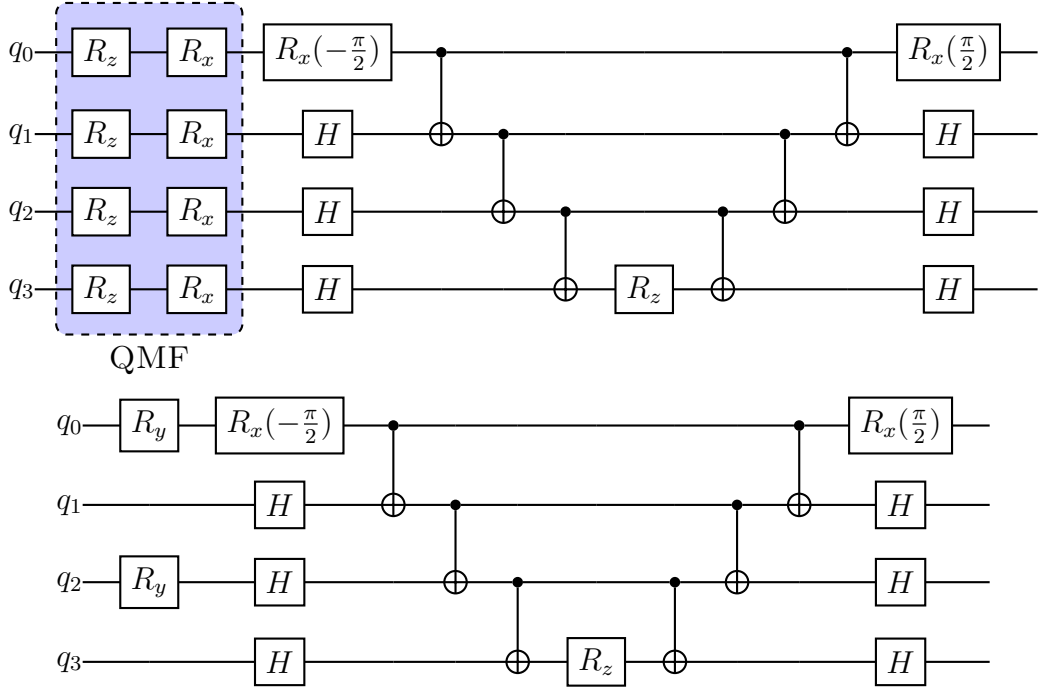

The quantum circuit is finally constructed using a ladder-like block procedure~\cite{barkoutsos2018quantum}, as shown in Fig.~\ref{fig:qcc}. As pointed out by Ref.~\cite{ryabinkin2018qubit}, for molecules like LiH and H$_2$O, the two-qubit gate count is greatly reduced compared to that of the UCC ansatz by bypassing any explicit Fermionic construction of electronic excitations and thus saving as many quantum resources as possible.

In the original proposal of the QCC~\cite{ryabinkin2018qubit} method, only entanglers with more than one non-identity gate ($X, Y, Z$) were considered, and  the quantum circuit was started with a qubit mean-field (QMF) component, as shown in Fig.~\ref{fig:qcc}. This QMF component consists of single $R_x$ and $R_z$ rotations on each qubit, allowing for access to any point on the Bloch sphere. Although it remains to be investigated whether such an implementation suffers from the Barren plateau problem~\cite{mcclean2018barren}, the QMF component resembles the hardware-efficient ansatz~\cite{kandala2017hardware} and might pose optimization challenges~\cite{wang2021noise} due to the large number of required variational parameters ($2 N_q$). In addition, $R_x$ and $R_z$ rotations are likely redundant degrees of freedom, since often times both the Hamiltonian and wavefunction of many-body systems  of interest are real.

To solve the potential optimization challenges introduced by the QMF component, we propose a modification of the original QCC ansatz. We simply discard the QMF component, and consider all the possible entanglers when performing the screening operation, regardless of the number of qubits that the entangler involves. In this way, the number of necessary variational parameters are reduced and eventually a quantum circuit only contains exponentials of entanglers, as shown in the bottom panel of Fig.~\ref{fig:qcc}.

The QCC ansatz is suitable for NISQ devices, where a trade off between circuit depth and number of quantum measurements is desirable. We note that similar ideas to construct efficient ansatz circuits using gradient methods have been explored in recent years, including iterative QCC~\cite{ryabinkin2020iterative}, ADAPT-VQE and its several variants~\cite{grimsley2019adaptive, tang2021qubit}, e.g. ClusterVQE~\cite{zhang2021variational}, factorized-form of UCC~\cite{chen2021quantum}, and projective quantum eigensolver~\cite{stair2021simulating}. Besides being hardware friendly, an additional benefit of the QCC ansatz is that it leads to differentiable potential energy surfaces given its functional dependence on the entanglers, and the gradients can be estimated using the parameter-shift rule~\cite{crooks2019gradients, schuld2019evaluating}. The QCC ansatz has shown the correct size-consistent behavior when applied to study the dissociation of H$_2$, LiH and H$_2$O in Ref~\cite{ryabinkin2018qubit}. However, there is no guarantee that it will always yield the correct behavior for any systems since it depends on how the entanglers are truncated when the circuit is constructed. We also speculate that in general size-extensivity may be satisfied as entanglers for noninteracting fragments act only on each fragment, and, therefore, commute and we thus have $E(2A)=2E(A)$.

\subsection{Quantum Subspace Expansion for Excitation Energies}

We now turn to the discussion of the calculations of excitation energies, for which subspace type methods~\cite{mcclean2017hybrid, colless2018computation, epperly2021theory, tkachenko2022quantum, cortes2022quantum, kirby2022exact} are suitable. These methods can be viewed as a quantum analog of CI and its variants, e.g., selected CI approach~\cite{buenker1974individualized}. Here we choose the quantum subspace expansion (QSE) algorithm~\cite{mcclean2017hybrid, colless2018computation, mcardle2020quantum}, which uses the same quantum circuit as the one to obtain the ground state and involves only additional quantum measurements~\cite{colless2018computation}. Specifically for the ground state $\ket{\Psi}$, a set of expansion operators $\{\hat{O}_i\}$ is chosen, which act on $\ket{\Psi}$ to form a basis given by $\{\hat{O}_i \ket{\Psi}\}$, where $\hat{O} \in \{\hat{a}^{\dagger}_a \hat{a}_i,\; \hat{a}^{\dagger}_a \hat{a}^{\dagger}_b \hat{a}_j \hat{a}_i | i,j\in\mathcal{A};\; a,b \in \mathcal{V}\}$. We use this basis to evaluate the Hamiltonian and overlap matrix elements:
\begin{equation}
    H^{\mathrm{QSE}}_{ij} = \bra{\Psi} \hat{O}^{\dagger}_i \hat{H} \hat{O}_j \ket{\Psi},\;\; S^{\mathrm{QSE}}_{ij} = \bra{\Psi} \hat{O}^{\dagger}_i \hat{O}_j \ket{\Psi}.
\end{equation}
Note that the expansion operators are not limited to double excitations, and we did not include additional excitations as double ones are sufficient to obtain the FCI spectrum of our systems. Using the matrices defined above, we then solve the generalized eigenvalue problem in the well conditioned subspace given by $H^{\mathrm{QSE}} C = S^{\mathrm{QSE}} C \varepsilon,$ where $C$ is the matrix of eigenvectors and $\varepsilon$ the diagonal matrix of eigenvalues. As mentioned in section~\ref{encoding}, the QEE encoding is used to transform the excitation operators $\hat{a}_i^{\dagger}\hat{a}_j$ into Pauli strings acting on $N_q$ qubits, and the matrix elements are evaluated as weighted sums of the expectation values of these Pauli strings. The cost of QSE has two components: i) determining the matrix elements through measurements, and ii) solving the generalized eigenvalue problem. In the QEE-QCC scheme adopted in this work, the measurement cost is negligible since the majority of Pauli operators have been measured already when computing the ground state. Therefore the cost of the QSE calculations mainly comes from ii). We also note that the effectiveness of QSE is achieved with a careful choice of creation and annihilation operators, which would be facilitated by using chemical intuition, e.g., by identifying the most dominant excitations.

\clearpage

\section{Results} \label{Results}

In this section we present results for the many-body ground and excited states of the \ce{NV-} center in diamond, \ce{VV^0} and \ce{V^-_{Si}} in 4H-SiC. Using the methods described in Sec.~\ref{Methods}, we performed calculations on the \textit{ibmq\_guadalupe} quantum computer using the IBM Qiskit package~\cite{aleksandrowicz2019qiskit}. We have applied measurement error mitigation~\cite{dewes2012characterization, maciejewski2020mitigation} to all the measurements.

\subsection{Reference results on classical hardware} \label{results:classical}

We use QDET to obtain the effective second quantized Hamiltonian, which is then used as input for our quantum computations~\cite{ma2021quantum, sheng2022green, vorwerk2022quantum}. As a first step we define a periodic supercell with hundreds of atoms, representing a crystal with a defect center embedded in it, and we compute its electronic structure using Kohn-Sham (KS) density functional theory (DFT) with the PBE functional, the $G_0W_0$ approximation, and the Quantum Espresso~\cite{giannozzi2009quantum, giannozzi2017advanced} and WEST~\cite{govoni2015large} codes. A subset of KS orbitals localized around the defect is then chosen based on the localization criterion defined in Ref.~\cite{sheng2022green}. This subset constitutes the so-called active space $A$ spanned by the second quantized effective Fermionic Hamiltonian $\hat{H}_{\text{elec}} = \sum_{ij}^A t_{ij}^{\mathrm{eff}} \hat{a}^{\dagger}_i \hat{a}_j + \frac{1}{2} \sum_{ijkl}^A v_{ijkl}^{\mathrm{eff}} \hat{a}^{\dagger}_i \hat{a}^{\dagger}_j \hat{a}_l \hat{a}_k $. The effective two-body matrix elements $v_{ijkl}^{\mathrm{eff}}$ are computed using the constrained random-phase approximation (cRPA) method. The effective one-body matrix elements  $t_{ij}^{\mathrm{eff}}$ are computed from the $G_0W_0$ Hamiltonian removing a double counting term. Notably, in Ref.~\cite{sheng2022green} we rigorously derived an expression of the double counting term within the $G_0W_0$ approximation.

\begin{figure}[t!]
    \centering
    \includegraphics[width=\textwidth]{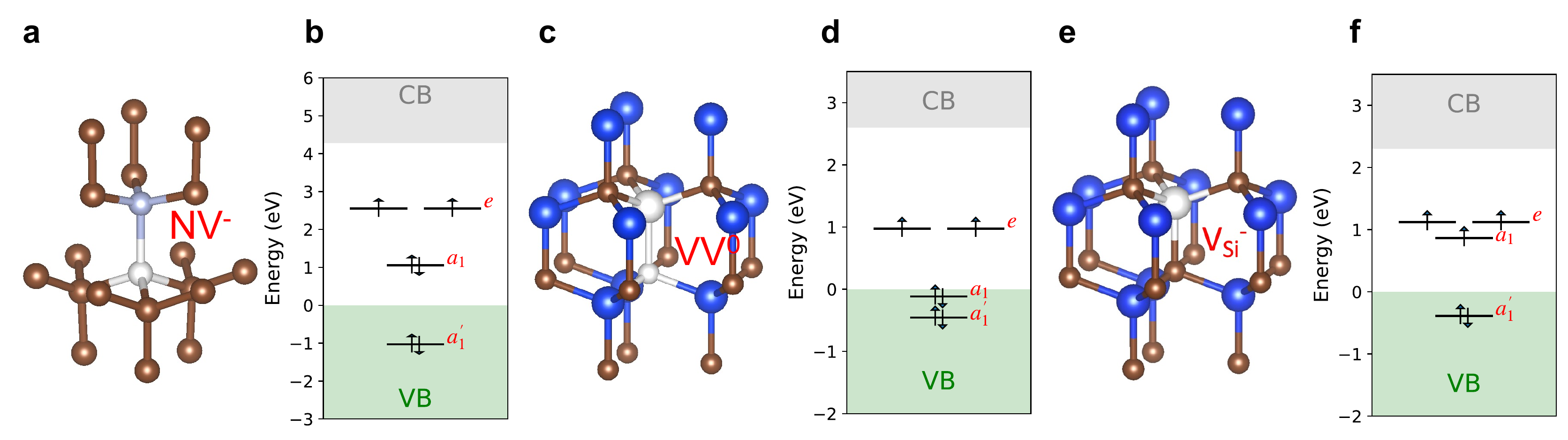}
    \caption{Spin defects studied in this work: the \ce{NV^-} center in diamond, the \ce{VV^0} and \ce{V^-_{Si}} in 4H-SiC. Panels \textbf{a}, \textbf{c} and \textbf{e} show a ball-and-stick representation of the defects. Panels \textbf{b}, \textbf{d} and \textbf{e} show single particle states obtained by solving the Kohn-Sham equations for the entire periodic solid, where gray and green shaded areas represent the conduction (CB) and valence band (VB), respectively; the single particles states are shown as black lines.}
    \label{fig:dft}
\end{figure}

\begin{figure}[t!]
    \centering
    \includegraphics[width=\textwidth]{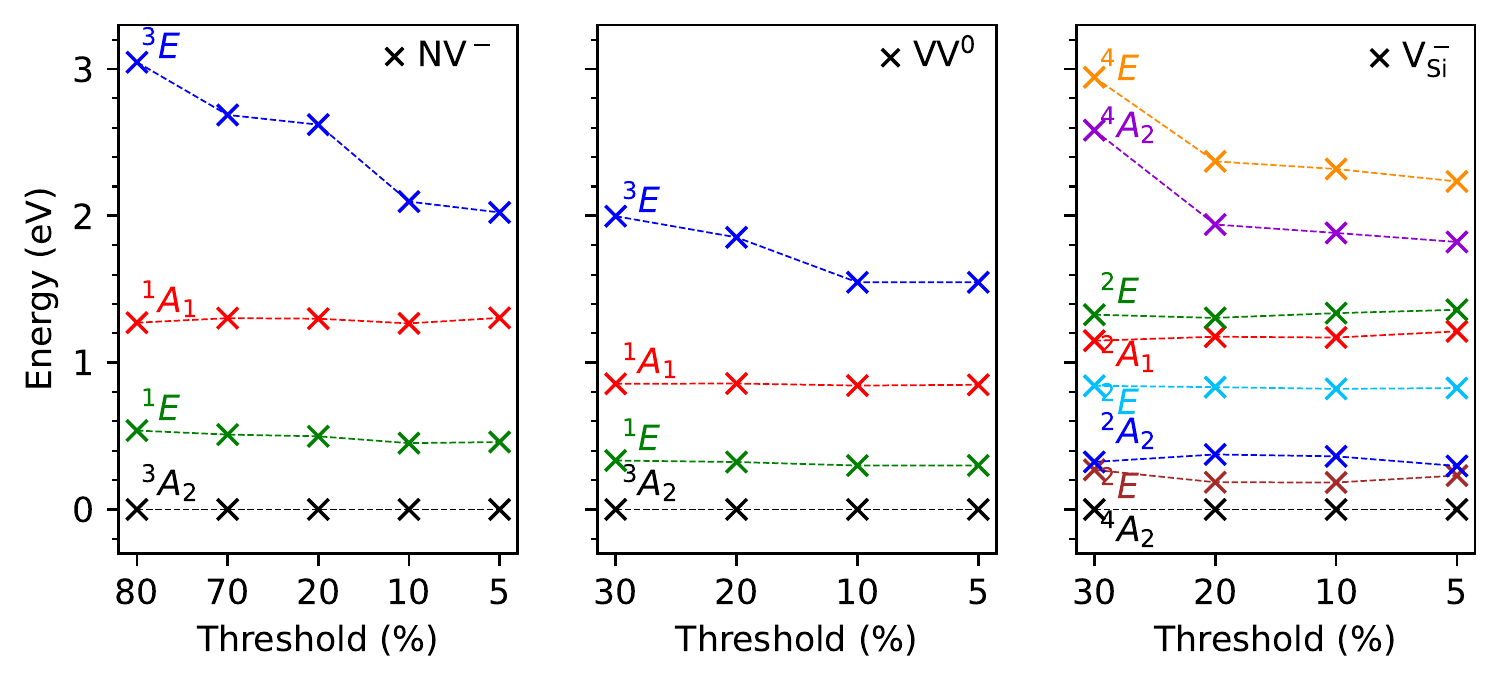}
    \caption{The left, middle and right panel show computed vertical excitation energies for the \ce{NV^-} center in diamond, \ce{VV^0} and \ce{V^-_{Si}} in 4H-SiC as a function of the chosen localization threshold. States are labeled using the irreducible representation of the $C_{3v}$ point group. We note that the largest threshold corresponds to a (4e, 3o), (8e, 5o) or (5e, 4o) active space for the three defects, respectively, and the smallest threshold corresponds to a (26e, 14o), (64e, 33o) or (57e, 30o) active space, respectively.}
    \label{fig:convergence_pbe}
\end{figure}

We computed the electronic structure of \ce{NV-}, \ce{VV^0} and \ce{V^-_{Si}} using a 215-, 198- and 127-atom supercell, respectively. We performed restricted closed-shell plane wave DFT calculations with the optimized structure from unrestricted open-shell calculations. We used the PBE~\cite{perdew1996generalized} exchange-correlation functional, SG15 norm-conserving pseudopotentials~\cite{schlipf2015optimization}, and a 50 Ry kinetic energy cutoff for the plane wave basis set. The active space was defined considering all KS orbitals with highest localization factor $L_V(\psi_n^{\textbf{KS}}) = \int_{V} |\psi_n^{\textbf{KS}}(\textbf{x})|^2 d\textbf{x}$, where the integration is performed on a predefined volume $V$ around the defect center (see Fig.~\ref{fig:convergence_pbe}), as originally defined in Ref.~\cite{sheng2022green}. In our full-frequency $G_0W_0$ calculations we used 512 projective dielectric eigenpotentials (PDEPs) to represent the dielectric response. The QDET method is implemented in the WEST code~\cite{govoni2015large, yu2022gpu}.

In Fig.~\ref{fig:convergence_pbe} we show the convergence of vertical excitation energies of the three defects w.r.t. the localization threshold, which sets a lower bound for the KS orbitals to be included in the active spaces. We note that the energies are relatively well converged at 10\%, 10\% and 20\% threshold in the three cases, corresponding to (14e, 8o), (22e, 12o) and (9e, 6o) active spaces, respectively. Due to the limitation in quantum resources, a compromise had to be made in selecting the active space to generate the effective Hamiltonians for the three defects: we chose the (14e, 8o) active space for \ce{VV^0} and \ce{NV^-}. For \ce{NV^-}, the convergence threshold for the (14e, 8o) active space lies between $20\%$ and $10\%$ and its excitation energies differ by approximately 0.1 eV from those obtained with a 10\% localization threshold. Considering the complexity of the \ce{V^-_{Si}}, for this defect  we had to resort to a 30\% localization threshold, leading to a (5e, 4o) active space, which we refer to as the ``minimum model''.

We also note that when the localization threshold is lowered, the size of the active space $N$ is increased by the inclusion of additional occupied orbitals. The total number of Slater determinants with a constant number of holes in each spin channel, $c_{\uparrow(\downarrow)} = \frac{N}{2} - m_{\uparrow(\downarrow)}$, scales polynomially as $O(N^{(c_{\uparrow}+c_{\downarrow})}$. This leads to an encoded effective Hamiltonian which is expressed as a linear combination of up to $O(N^{2(c_{\uparrow}+c_{\downarrow})}$ Pauli operators. The size of the QEE Hamiltonian and its corresponding classical preprocessing step do not pose a computational challenge for the  systems considered here, where $c_{\uparrow(\downarrow)} \leq 2$.

\subsection{Calculation of the ground state using a quantum computer}

\subsubsection{\texorpdfstring{\ce{VV^0}}{} in 4H-SiC and \texorpdfstring{\ce{NV-}}{} in diamond}

The ground state of the effective Hamiltonian constructed for both the \ce{VV^0} in 4H-SiC and the \ce{NV-} center in diamond is a ${}^3A_2$ triplet state, whose $m_S=0$ state has a multi-reference character. To obtain such a state on quantum computers using VQE, a good guess for the initial wavefunction is key to achieving fast convergence~\cite{huang2022simulating}. These systems are open-shell with the highest occupied molecular orbitals (HOMOs) in the active space being $e$ orbitals; hence it is wise to use $|\Psi_0\rangle = |..a_1 \overline{a}_1 e_x \overline{e}_y\rangle$ due to Hund's Rule, where $a_1$, $e_x$, $e_y$ (spin-up) and $\bar{a}_1$, $\bar{e}_x$, $\bar{e}_y$ (spin-down) denote the single particle orbitals in the active space, as shown in Fig.~\ref{fig:dft}.

As mentioned above, we choose the (14e, 8o) active space for both defects, leading to a total of 64 Slater determinants, and requiring the use of 6 qubits to span the full qubit Hilbert space, i.e.,  $\mathcal{H}_{\mathrm{eff}} \subseteq \mathcal{H}_{\mathrm{q}}$. To construct the QCC ansatz, we first measure the energy gradients of different entanglers using Eq.~\ref{eq:first_order}. The amplitude of these gradients are obtained both on a real quantum device (noisy) and on a simulator (noiseless). They are listed in Tab.~\ref{tab:prescreening} for \ce{VV^0}. We find that the difference between noisy and noiseless results is negligible (within $1\sim 2\%$). This is due to the fact that the measurement circuits does not contain two-qubit gates, which are major sources of error in NISQ devices. We also note that the four entanglers with top rank correspond to entanglers with the identity gate $I$ at the two left-most qubit indices. Because of the chosen QEE scheme, the two left-most qubit indices control Slater determinants with the highest excitation energy, and originate from transitions from the lowest four occupied single particle orbitals. This suggests that we can use the frozen core approximation~\cite{rossmannek2021quantum} to reduce the computational cost without sacrificing accuracy. Therefore we freeze the lowest four occupied orbital and in practice we work with 4 qubits, 4 entanglers, and a qubit Hamiltonian with 136 terms. The QCC circuit is constructed using 14 CNOT gates in total. The UCC counterpart, however, would require $\sim 400$ CNOT gates, indicating the critical advantage of the QCC method. For \ce{NV-}, the same logic applies and its circuit has 10 CNOT gates. Note that in the interest of generality, during the construction of the QCC ansatz, we did not take into consideration the point group symmetry of the ground state so as to design calculations that would be viable also for systems under strain or for moderately disordered lattices. If symmetry is invoked, both UCC and QCC can be reduced to a simple circuit with a single parameter and only two CNOT gates~\cite{huang2022simulating}.

\begin{table}
\caption{\label{tab:prescreening} Top entanglers for the electronic structure calculation of \ce{VV^0} and \ce{V^-_{Si}} from Eqn.~\ref{eq:first_order} before the frozen core approximation is carried out, with their magnitude computed using a noiseless simulator and a quantum hardware \textit{ibmq\_guadalupe} (Atomic units)}
\centering
\begin{tabular}{|*{7}{c|}}
\hline
Rank & \multicolumn{3}{c|}{\ce{VV^0}} & \multicolumn{3}{c|}{\ce{V^-_{Si}}}\\
\cline{2-7}
& Entanglers & Noiseless & Noisy & Entanglers & Noiseless & Noisy \\
\hline
1 & $IIIIXY$ & 0.009243 & 0.009170 & $IXYII$ & 0.006969 & 0.006754\\
2 & $IIXIYZ$ & 0.008177 & 0.008100 & $IIIYI$ & 0.006693 & 0.006601\\
3 & $IIXXIY$ & 0.008165 & 0.008065 & $IIIIY$ & 0.004352 & 0.004352\\
4 & $IIXIXY$ & 0.006587 & 0.006529 & $IIIXY$ & 0.004350 & 0.004350\\
\hline
\end{tabular}
\end{table}

Results from the VQE optimization of \ce{VV^0} and \ce{NV^-} are shown in the upper and middle panel of Fig.~\ref{fig:vqe}. The energy is evaluated on the quantum hardware as the weighted sum of the expectation values of Pauli strings, i.e., $E=\sum_i g_i \langle\hat{P}_i\rangle$. The expectation values of all Pauli strings were obtained by measuring 8192 times $N_c$ independent circuits so that the standard deviation ($\sigma$) of measurement is within 15 meV, where $N_c$ is the number of groups that contain mutually commuting strings. In the case of \ce{VV^0} (\ce{NV^-}), we find that the VQE calculation converges to a state that is $\sim0.5$ ($0.4$) eV higher than the FCI reference energy obtained on a classical computer. The fluctuations are more pronounced for the \ce{NV^-} center because our calculations were carried out at different times and the hardware environment was not identical for each measurement. The insets in Fig.~\ref{fig:vqe}, where the reference values are from noiseless simulations, show that in our optimization procedure  we indeed converge to the ground state of the system. We find that for both \ce{VV^0} and \ce{NV^-}, only the parameter associated with entangler $IIXY$ is nonzero ($\pi/2$), indicating that the other ones are negligible in determining the ground state, thus reducing the circuit to one exponential block of $IIXY$. This simplified circuit is exactly what we obtained in Ref.~\cite{huang2022simulating} by taking into consideration the point group symmetry of the lattice.

To obtain an accurate estimate of the ground state energy, error mitigation is required and here we adopted the ZNE method. The latter is straightforward to implement and does not require additional qubits. The basic idea of ZNE is to amplify the noise of the circuit to various controllable levels and obtain the zero noise limit by extrapolation. The key to success of ZNE lies in how noise is artificially boosted. We employ a split exponential technique that we originally proposed in Ref.~\cite{huang2022simulating} to artificially increase the circuit depth of each exponential block $e^{i\theta_k \hat{P}_k}$ of the QCC quantum ansatz, i.e., $n$ replicas are generated with $\left(e^{i\frac{\theta_k}{n} \hat{P}_k}\right)^n$. We note that this technique is suitable for both UCC and QCC-type of ansatzes~\cite{lee2018generalized, metcalf2020resource, liu2021unitary, fedorov2022unitary, ryabinkin2018qubit} and it does not affect the Trotter error~\cite{huang2022simulating}. The extrapolation procedure is shown in Fig.~\ref{fig:extrapolation}. We worked with the reduced circuit with only one entangler $IIXY$, and considered $n=[1,2,3,4,5]$. For each value of $n$, we increased the measurements to 320000, so $\sigma$ is kept within 2.5 meV and the stability of the extrapolation procedure is improved. A quadratic function is used for extrapolation and the difference between the ground state energy and the reference value obtained on a quantum simulator is one order of magnitude smaller than in the absence of ZNE.


\subsubsection{\texorpdfstring{\ce{V^{-}_{Si}}}{} in 4H-SiC}

We now turn to the discussion of the V$_{\mathrm{Si}}^-$ spin defect in 4H-SiC, which has a $\ket{{}^4A_2}$ ground state, with multi-reference character for $m_s = \pm \frac{1}{2}$~\cite{soykal2016silicon}. The HOMOs of V$_{\mathrm{Si}}^-$ consists of three quasi-degenerate orbitals: $a_1, e_x, e_y$ that are all singly occupied in the ground state, as shown in Fig.~\ref{fig:dft}. Therefore the $m_s = \pm \frac{1}{2}$ spin manifold is considerably more complicated than those of \ce{VV^0} and \ce{NV-}. We use only a minimal model of (5e, 4o) for the active space to describe this system, which is adequate to demonstrate the advantages of QCC over UCC in terms of finding the ground state with a shallow circuit depth. In this minimal model, the $m_s = \frac{1}{2}$ component of the ground state wavefunction consists of 6 Slater determinants
\begin{equation}
\begin{split}
    \ket{\Psi_g} & = \alpha \left(\ket{a_1^{\prime} \overline{a_1}^{\prime} \overline{a_1} e_x e_y} + \ket{a_1^{\prime} \overline{a_1}^{\prime} a_1 \overline{e_x} e_y} + \ket{a_1^{\prime} \overline{a_1}^{\prime} a_1 e_x \overline{e_y}}\right)\\
    & + \beta \left(\ket{\overline{a_1}^{\prime} a_1 \overline{a_1} e_x e_y} + \ket{a_1^{\prime} a_1 \overline{a_1} \overline{e_x} e_y} + \ket{a_1^{\prime} a_1 \overline{a_1} e_x \overline{e_y}}\right), \label{V_si}
\end{split}
\end{equation}
where we have only used two coefficients, $\alpha$ and $\beta$ because of symmetry. From the FCI solutions on a classical computer we know that the first three configurations with doubly occupied $a_1^\prime$ are dominant ($|\alpha|=0.576,\;|\beta|=0.0391$), hence we use one of them as the initial state of our VQE optimization. Specifically, we use $\ket{\Psi_0} = \ket{a_1^{\prime} \overline{a_1}^{\prime} \overline{a_1} e_x e_y}$.

\begin{figure}[t!]
    \centering
    \includegraphics[width=0.8\textwidth]{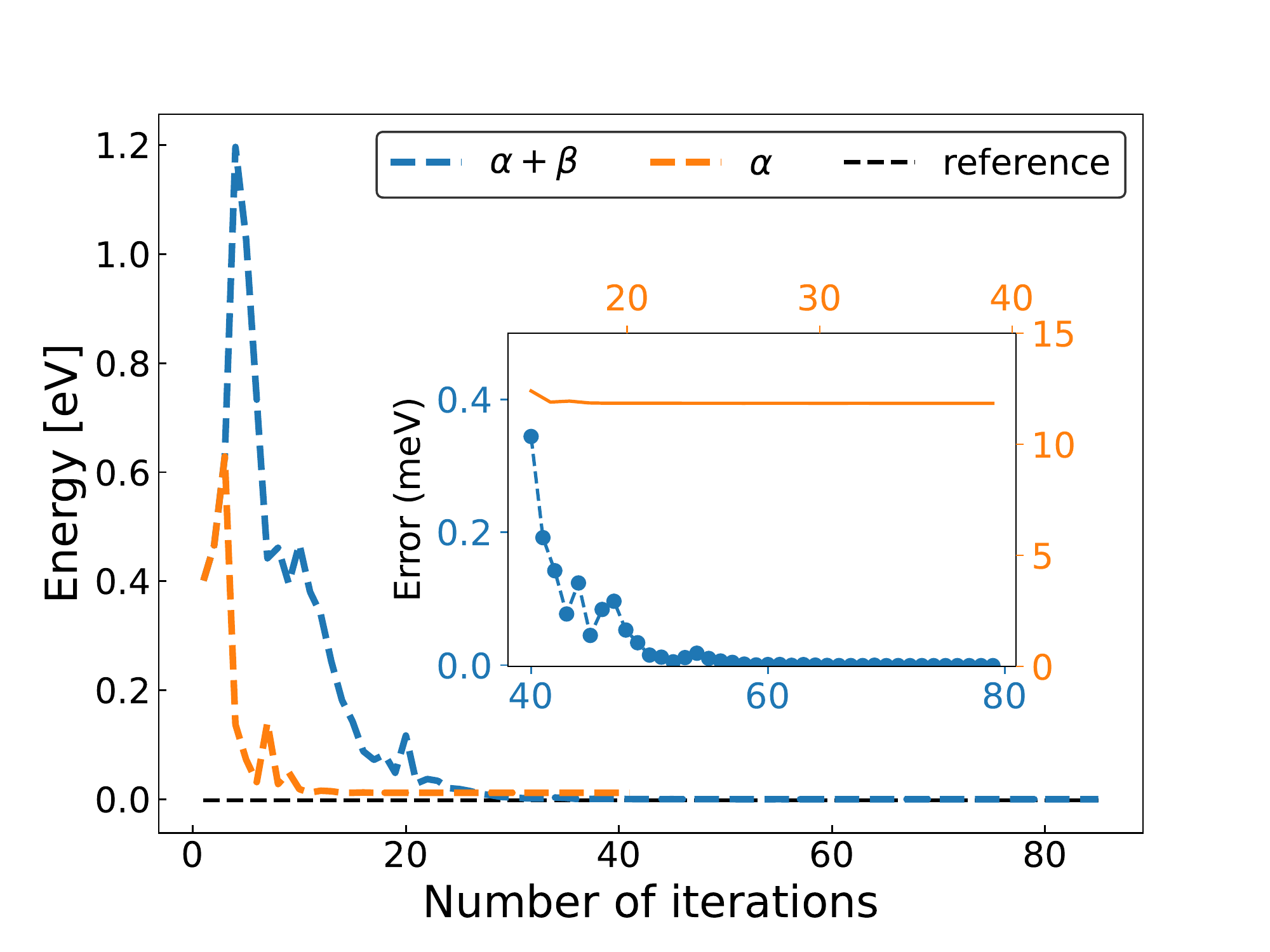}
    \caption{Total energy as a function of the number of iterations used to optimize the ground state energy of \ce{V^-_{Si}} in 4H-SiC using the variational quantum eigensolver (VQE) algorithm on a noiseless simulator, with a unitary coupled cluster (UCC) ansatz and the COByLA optimizer~\cite{powell1994direct}. The blue and orange curves represent results using two variants of the ansatz circuit with different levels of approximation; see text. The inset shows the error of different VQE optimizations relative to the reference energy. The full configuration interaction (FCI) energy (dashed black line) is reported for reference.}
    \label{fig:vsi-ucc}
\end{figure}

When adopting the UCCSD ansatz, one needs to explicitly construct the relevant electronic excitations, whose associated parameters are $\theta_{a_1^\prime}^{a_1}, \theta_{e_x \overline{a_1}}^{a_1 \overline{e_x}}, \theta_{e_y \overline{a_1}}^{a_1 \overline{e_y}}, \theta_{e_x \overline{a_1}^\prime}^{a_1 \overline{e_x}}, \theta_{e_y \overline{a_1}^\prime}^{a_1 \overline{e_y}}$. The resulting circuit requires $\sim$200 CNOT gates, and with the UCCSD ansatz we only obtained the exact ground state on a noiseless simulator. A reasonable approximation is to assume $\beta=0$, which would require only the first two parameters with a corresponding reduction of the number of CNOT gates to $\sim$80. This approximation leads to an error  of $\sim$11 meV, as shown in~\ref{fig:vsi-ucc}. However, both circuits are beyond the capability of NISQ quantum devices.

\begin{figure}[t!]
    \centering
    \includegraphics[width=0.75\textwidth]{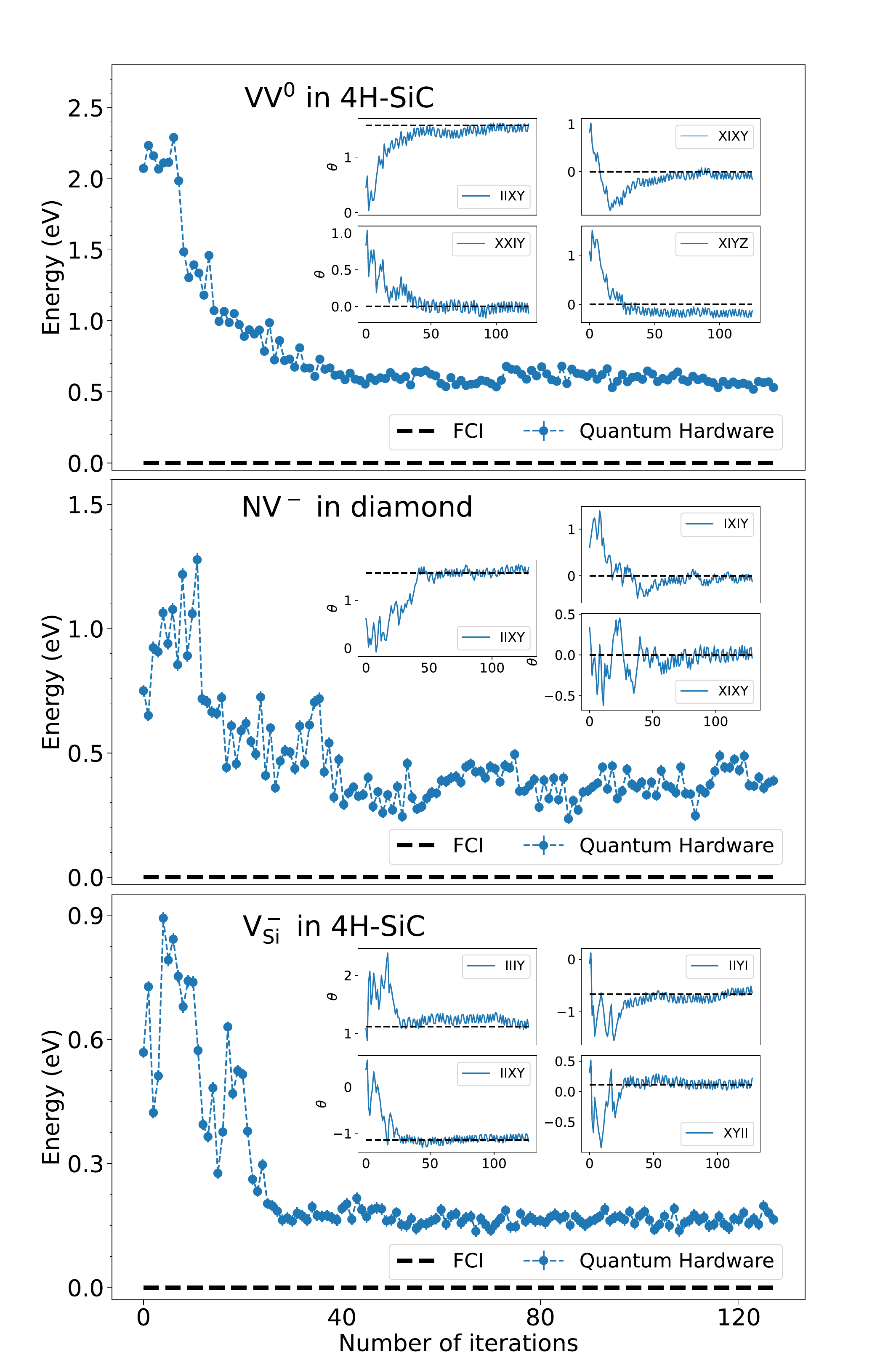}
    \caption{The upper, middle and bottom panel show the total energy as a function of the number of iterations during an optimization of the ground state energy of the \ce{VV^0} in 4H-SiC, the \ce{NV^-} in diamond and the \ce{V^-_{Si}} in 4H-SiC carried out with the variational quantum eigensolver (VQE) algorithm on \textit{ibmq\_guadalupe} (quantum hardware); the variation of parameters associated with each entangler of the qubit coupled cluster (QCC) ansatz is plotted in the inset. The full configuration interaction (FCI) energy is reported for reference.}
    \label{fig:vqe}
\end{figure}

\begin{figure}[t!]
    \centering
    \includegraphics[width=0.8\textwidth]{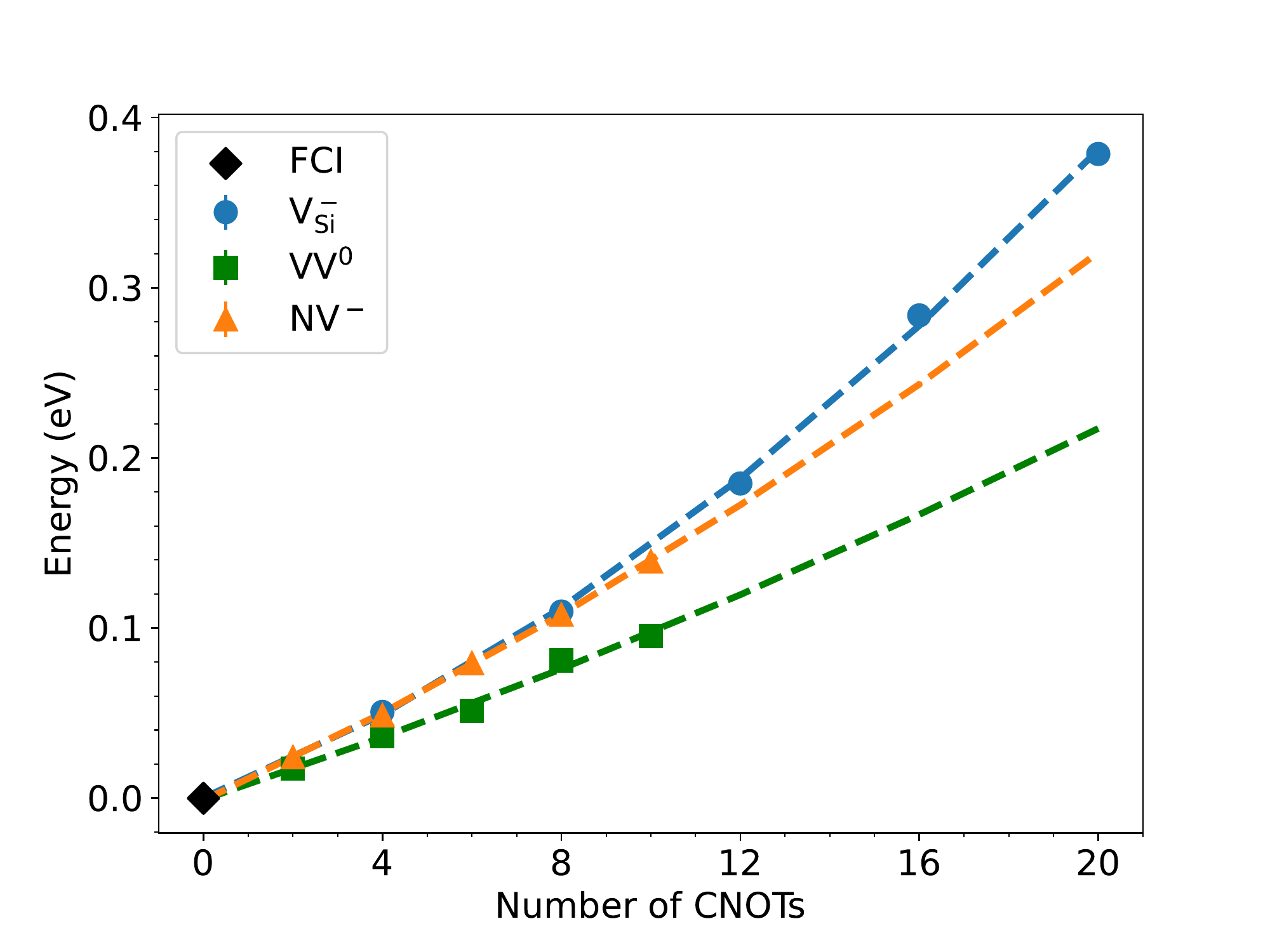}
    \caption{The ground state energy of the \ce{NV^-} center and the \ce{VV^0} and \ce{V^-_{Si}} in 4H-SiC as a function of the number of replicas used in the zero-noise extrapolation (see text), obtained using \textit{ibmq\_guadalupe}. The $x$ axis is scaled with the number of CNOT gates used in the quantum circuit for clarity of comparison. The reference, noiseless result has been set to 0.}
    \label{fig:extrapolation}
\end{figure}

Here the QCC ansatz presents a remarkable advantage and we were able to simulate the \ce{V^-_{Si}} defects  on a real quantum processor. The screening of entanglers are summarized in Tab.~\ref{tab:prescreening}. We selected the entanglers with top rank to construct the circuit for the ansatz, which contains only a total of 4 CNOT gates. The VQE optimization on a real quantum processor is shown in Fig.~\ref{fig:vqe}, where the error due to noise is about $\sim$0.2 eV. We note that also in this case the ZNE is applied at the end of the VQE optimization to obtain a more accurate ground state energy, as shown in Fig.~\ref{fig:extrapolation}.

\subsection{Calculation of the excited states using a quantum computer} \label{QSE}

As mentioned earlier, we computed excited states of the \ce{VV^0} and \ce{NV^-} using the QSE algorithm. To avoid propagating the errors introduced by VQE, we used the exact energy of the ${}^3A_2$ state with $m_s=0$ as the ground state energy.

We constructed a quantum subspace that is identical to the configuration state space, so the dimension of the QSE matrices is the same as that of their classical FCI counterpart. The QSE matrix is built by evaluating, on the quantum hardware, the expectation values of all Pauli strings. In our zero noise mitigation, we used a linear extrapolation for the off-diagonal elements of the QSE matrix and we computed diagonal elements with linear and quadratic extrapolations. The number of measurements was 320000 for both defects. The QSE matrix was finally diagonalized on a classical computer.

\begin{figure}[t!]
    \centering
    \includegraphics[width=0.8\textwidth]{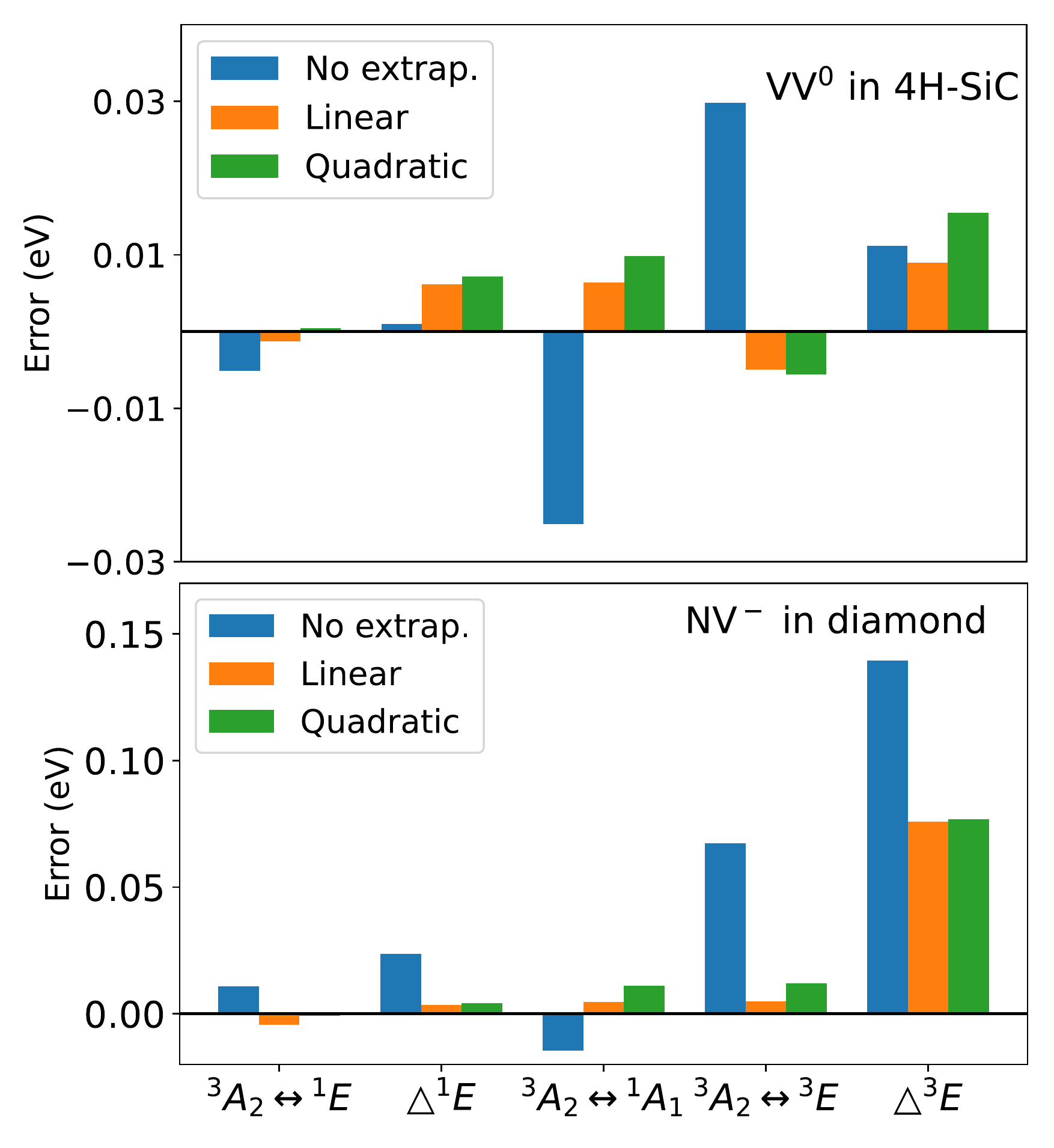}
    \caption{The upper and bottom panel show the error in the excitation energies (eV) of the \ce{VV^0} and \ce{NV^-} defects calculated using the quantum subspace expansion (QSE) method on \textit{ibmq\_guadalupe}. The $x$ axis shows transitions between states labeled using the representation of the point group $C_{3v}$, following Ref.~\cite{ma2020quantum}. $\Delta ^1E$ and $\Delta ^3E$ indicate the breaking of degeneracy due to noise (see text).  The reference values are obtained with a noiseless simulator and are identical to those of classical full configuration interaction (FCI) calculations on a classical computer. The blue, orange and green bar represent results obtained using no extrapolation, linear and quadratic zero-noise extrapolation techniques, respectively. For the results labeled with “quadratic extrapolation”, we only carried out a quadratic extrapolation for the diagonal elements of the QSE matrix elements, and a linear extrapolation was applied to the off-diagonal elements.}
    \label{fig:qse}
\end{figure}

The errors of excitation energies with and without extrapolation are summarized in Fig.~\ref{fig:qse}. The accuracy of the energy of non-degenerate excitations is in general improved when using the ZNE. We note that overall different choices of extrapolation functions lead to similar results, and hence linear extrapolation is a desirable choice, since a smaller number of parameters is expected to lead to a more stable fit. The degeneracy of states is spuriously lifted on the quantum hardware due to the presence of noise, though it is slightly mitigated after applying the linear ZNE.

\clearpage

\section{Conclusions} \label{Conclusions}

In summary, we presented a computational protocol to diagonalize Fermionic Hamiltonians on noisy-intermediate-quantum computers, which combines the QEE scheme to map electronic excitations onto qubits, a modified QCC ansatz for VQE optimizations of the ground state and noise mitigation techniques. The QEE mapping offers a robust solution to the unphysical state problem and the QCC ansatz provides a relatively short quantum circuit suitable for calculations on near-term intermediate-size quantum devices. We applied our protocol on quantum hardware to compute the electronic structure of strongly correlated ground and excited states of three spin defects, i.e., the \ce{NV^-} center in diamond, the \ce{VV^0} and \ce{V^-_{Si}} in 4H-SiC, and we presented calculations that would have been unfeasible with conventional algorithms. In particular, we could go beyond the minimum models for the \ce{NV^-} and \ce{VV^0} and tackle a complex defects such as \ce{V^-_{Si}} for the first time. Work is in progress to improve the efficiency of the measurements of $\langle H\rangle$ on quantum architectures, for example by adopting advanced measurements techniques with different term groupings~\cite{gui2020term, yen2022deterministic}, fragmentation procedures~\cite{choi2022improving, choi2022fluid} and classical shadow~\cite{huang2020predicting, nakaji2022measurement}, and to extend the applicability of our protocol to larger active spaces appropriate, e.g. to investigate adsorbates on surfaces or ions and nanostructures in solution. We finally note that establishing which algorithms are better suited to achieve quantum advantage in electronic structure calculations remains an open area of research. For example, recent papers have argued that simulations in first quantization offer some important advantages over approaches in second quantization including faster convergence to the continuum limit and the opportunity for practical simulations beyond the Born-Oppenheimer approximation~\cite{su2021fault}. Interestingly, in addition to efforts towards reaching a practical advantage with quantum computers, the development of algorithms for quantum computations is having a positive impact on the development of classical algorithms in various fields, e.g., machine learning~\cite{tang2019quantum} and computational spectroscopy~\cite{oh2022quantum}.

\clearpage

\section*{Acknowledgments}

We thank Yu Jin for many fruitful discussions. We also thank the Qiskit Slack channel for generous help. This work was supported by the computational materials science center Midwest Integrated Center for Computational Materials (MICCoM) for the implementation and use of quantum embedding and by the Next Generation Quantum Science and Engineering (QNEXT) hub for the development of quantum algorithms and deployment on quantum hardware. MICCoM is part of the Computational Materials Sciences Program funded by the U.S. Department of Energy, Office of Science, Basic Energy Sciences, Materials Sciences, and Engineering Division through the Argonne National Laboratory, under Contract No. DE-AC02-06CH11357. QNEXT is supported by the U.S. Department of Energy, Office of Science, National Quantum Information Science Research Centers. This research used resources of the Oak Ridge Leadership Computing Facility at the Oak Ridge National Laboratory, which is supported by the Office of Science of the U.S. Department of Energy under Contract No. DE-AC05-00OR22725, resources of the National Energy Research Scientific Computing Center (NERSC), a DOE Office of Science User Facility supported by the Office of Science of the U.S. Department of Energy under Contract No. DE-AC02-05CH11231, and resources of the Argonne Leadership Computing Facility, which is a DOE Office of Science User Facility supported under Contract No. DE-AC02-06CH11357. We acknowledge the use of IBM Quantum services for this work and to advanced services provided by the IBM Quantum Researchers Program. The views expressed are those of the authors and do not reflect the official policy or position of IBM or the IBM Quantum team.

\section*{Author Contributions}
B.H., M.G., and G.G. designed the research. N.S. performed the embedding calculations on classical computers. B.H. conducted the experiments on IBM quantum computers, with supervision by M.G. and G.G. All authors contributed to the writing of the  manuscript.

\section*{Notes}
The authors declare no competing interests.

\bibliography{bibliography}

\end{document}